%Paper: hep-ph/9311301
%From: HEGYI@rmk530.rmki.kfki.hu
%Date: Tue, 16 Nov 1993 13:06 GMT+1

\magnification=\magstep1
\raggedbottom

\font\bo=cmbx10 scaled\magstep1
\font\tsc=cmcsc10

\hsize=15.truecm  \hoffset=.5truecm
\vsize=23.truecm  \voffset=.5truecm
\topskip=1.truecm \leftskip=0.truecm

\pageno=1
\footline={\hfill}
\headline={\ifnum\pageno=1\hfil\else\hss\tenrm-\ \folio\ -\hss\fi}

\let\cl\centerline  \let\rl\rightline   
\let\bs\bigskip    \let\ss\smallskip
\let\ub\underbar

\null
\rl{hep-ph/9311301}
\rl{November 1993}
\vskip2.truecm
{\bo
\cl{MONOFRACTAL DENSITY FLUCTUATIONS}\ss
\cl{AND SCALING LAWS FOR}\ss
\cl{COUNT PROBABILITIES AND COMBINANTS}
}

\vskip1.5truecm
\cl{\tsc S. Hegyi}

\footnote\ {\it hegyi@rmki.kfki.hu}

\baselineskip=11pt
{\it
\cl{KFKI Research Institute}
\cl{for Particle and Nuclear Physics}
\cl{of the Hungarian Academy of Sciences,}
\cl{H--1525 Budapest 114, P.O. Box 49. Hungary}
}
\bs

\vfill

\baselineskip=12pt
\midinsert\narrower\narrower
\noindent
{\rm
\underbar{ABSTRACT}\hskip.3truecm
The relation of
combinants to various statistics characterizing the fluctuation
pattern of multihadron final states is discussed.
Scaling laws are derived for count probabilities and combinants
in the presence of homogeneous and clustered mono\-fractal density
fluctuations. It is argued that both types of scaling rules are
well suited to signal Quark-Gluon Plasma formation in a second-order
QCD phase transition.
}
\endinsert
\vfill
\cl{To appear in Physics Letters B}
%cl{Working paper}
\vfill\eject

\baselineskip=13pt
\parskip=10pt

\noindent
{\bo 1. Introduction}

There is a renewed interest in an almost forgotten
statistic well suited to characterize the fluctuation pattern and
the nature of correlations of multihadron final states. This
statistic is the set of combinants introduced 15 years ago by
Gyulassy and Kauffmann~[1-2]. After their introduction the combinants
received little attention, only three papers have been published
concerning these quantities
\hbox{[3-5]}. Probably this was caused by the
fact that the existence of combinants requires nonzero probability
for detecting no particles at all (and precise data for this
probability) which was the exception rather than the rule 15 years ago
in the analysis of  multiplicity distributions. It is worth mentioning
that the Buras-expansion of multiplicity distributions invented years
earlier~[6] also utilizes the combinants but the emphasis in this work
was made on the factorial cumulant moments and the above important
requirement was unnoticed. This expansion was applied in~[7-8]. Three
further publications investigating the combinants have been appeared
quite recently~[9-11]. The work of Szapudi and Szalay~[10] is
particularly
recommended to the interested reader. Although not explicitly mentioned
the combinants are present also in the work of Lupia, Giovannini and
Ugoccioni~[12].

This letter concentrates on the
study of various scaling laws of count probabilities and combinants
that appear in the presence of monofractal density fluctuations.
The discussion of the relationship between combinants and the more
conventional measures of fluctuations and correlations is also the
goal of the paper. Section 2 summarizes the basic properties
of the combinants and a scaling law is provided which follows from the
validity of the Linked Pair Approximation (LPA).
A similar scaling law can be derived for count
probabilities if the observed events have monofractal structure. This
result will be
presented in section 3. In section 4 the two types of scaling rules
are reinvestigated for a Poisson superposition of clusters. Our
conclusions are in section 5.

\noindent
{\bo 2. Combinants, LPA and scaling}

The generating function ${\cal G}(z)$ of the multiplicity distribution
$P_n$ is conveniently defined by
$$
    {\cal G}(z)=\sum_{n=0}^\infty P_nz^n.\eqno(2.1)
$$
In the analysis of various count probability distributions occuring in
nature such as the distribution of galaxy-, hadron- and photocounts
the most successfully applied distribution functions are infinitely
divisible (e.g. Poisson, Gaussian, Negative Binomial, Lognormal).
For discrete distributions this family is known also as the compound
Poisson~[13]. A discrete distribution is said to be infinitely divisible
if its generating function
has the property that for all $k>0$ integer
$\root k\of {{\cal G}(z)}$
is again the generating function of a certain distribution~[13].
${\cal G}(z)$ satisfies this property if and only if
$
    {\cal G}(1)=1
$
and
$$
    \ln{\cal G}(z)=\ln{\cal G}(0)+\sum_{q=1}^\infty C_qz^q\eqno(2.2)
$$
where
$
    C_q\geq0
$ and
$$
    \sum_{q=1}^\infty C_q=-\ln{\cal G}(0)<\infty.\eqno(2.3)
$$
Hence the probability of detecting no particles, the so-called void
probability, is
$
    P_0={\cal G}(0)>0
$
for infinitely divisible distributions.
The quantities $C_q$ in eqs. \hbox{(2.2-3)} are the combinants
first studied in detail
by Gyulassy and Kauffmann \hbox{[1-2]}. We can get the $C_q$
by successively differentiating the logarithmic generating function:
$$
    C_q={1\over q!}{d^q\over dz^q}\ln{\cal G}(z)\,\Big|_{\,z=0}
    \eqno(2.4)
$$
and  ${\cal G}(z)$ takes the form
$$
    {\cal G}(z)=\exp\bigg(\sum_{q=1}^\infty C_q(z^q-1)\bigg)
    \eqno(2.5)
$$
for infinitely divisible distributions.
It is worth to introduce the count probability ratios
$
    {\cal P}_n=(P_n/P_0).
$
These can be expressed in terms of combinants by the recursion~[10]
$$
    {\cal P}_n={1\over n}\sum_{q=1}^n q\,C_q{\cal P}_{n-q}.
    \eqno(2.6)
$$
The ${\cal P}_n$ involve only a finite number of combinants but
$P_0$ (hence each $P_n$) requires all the $C_q$ as is seen from
eq.~(2.3). In terms of probabilities the combinants are found to be
$$
    C_q={\cal P}_q-{1\over q}\sum_{n=1}^{q-1} n\,C_n{\cal P}_{q-n}.
    \eqno(2.7)
$$
Essentially this expression was obtained also in ref.~[12] in a
different context.
{}From eq.~(2.7) two advantageous features of the combinants are
immediately seen. First, they require the knowledge of only a
{\it finite\/}
number of probabilities $P_n$. In $q$-th order the $P_{\leq q}$
are required.
Second, we need not know the probabilities themselves. The
combinants follow directly from the unnormalized topological cross
sections since they involve only {\it ratios\/} of probabilities.
It is also seen from eq.~(2.7) that in general
the combinants can take negative values as well for $q\geq2$.
In this case
a necessary condition of the infinite divisibility of
the multiplicity distribution $P_n$ is not satisfied.

In the analysis of correlations in restricted domains of phase-space
a basic set of quantities are the integrated cumulant
correlation functions providing the factorial cumulant moments,
$f_q$, of the underlying multiplicity distribution
\hbox{[14-15]}.
The factorial cumulants are defined by the Taylor expansion of
the logarithmic generating function:
$$
    \ln {\cal G}(z)=\sum_{q=1}^\infty{(z-1)^q\over q!}\,f_q.
    \eqno(2.8)
$$
One of the main advantages of the combinants is their close
relationship to the factorial cumulant moments.
This manifests itself e.g. in the very simple expressions
connecting the two quantities.
The factorial cumulants can be expressed in terms of
combinants according to
$$
    f_q=q!\sum_{n=q}^\infty{n\choose q}\,C_n\eqno(2.9)
$$
and the combinants read as
$$
    C_n=\sum_{q=n}^\infty{q\choose n}\,{(-1)^{q-n}\over q!}
    \,f_q\eqno(2.10)
$$
in terms of factorial cumulants~[2,6,10].
They also share some important common features. For the Poisson
distribution we have $C_1=\bar n$ and $C_{\geq2}=0$.
Hence nonvanishing higher-order combinants measure the degree of
deviation from Poissonian behaviour. The combinants exhibit
the additivity property of the cumulant moments. When a random
variable is composed of statistically independent random variables
the corresponding combinants are additive in those of the
independent components~[2].

In the last few years much interest has been devoted to the
analysis of correlations in the framework of the Linked Pair
Approximation [16-17]. In the LPA the  normalized
factorial cumulant moments,
$K_q=f_q/\bar n^q$, obey the recurrence relation
$$
    K_q=A_qK_2^{q-1}\eqno(2.11)
$$
where the coefficients $A_q$ should ideally be independent of
reaction type, energy, binsize and phase-space dimension.
In hadron-hadron collisions at $\sqrt s=200$ -- 900 GeV the validity of
LPA was confirmed with constant $A_{3,4}$ over the investigated
range of energy and binsize~[18]. The large uncertainties
of $A_5$ permit a clear conclusion.
At lower energies, $\sqrt s=22$ GeV, the constancy of the $A_q$
over binsizes was found to be violated by the NA22 Collaboration~[19].
This may be caused by the absence of translation invariance.
In nucleus-nucleus collisions the
$A_q$ are essentially zero indicating that only two-particle
correlations are present~[20].

Beyond the direct test of eq.~(2.11) using the factorial cumulants
there is another possibility
to check the LPA scheme. It is provided by a
scaling feature of the combinants~[11]. As is seen from eq.~(2.10)
if the $K_q$ obey the LPA relation
the ratios $C_n/\bar n$ chosen at fixed $n$
should scale to a function of $\bar nK_2$ only:
$$
    {C_n\over\bar n}=
    \sum_{q=n}^\infty{q\choose n}{(-1)^{q-n}\over q!}
    A_q\, (\bar nK_2)^{q-1}=\chi_n(\bar nK_2).\eqno(2.12)
$$
Increasing $n$ the contribution of an increasing number
of low-order factorial cumulant moments is excluded
from the $\chi_n$.
Eq.~(2.12) provides the generalization of
the scaling law obtained for the void probability
in the framework of the LPA [21,12].
As we shall see scaling combinants may appear also in the presence
of monofractal density fluctuations.

\noindent
{\bo 3. Monofractals and scaling count probabilities}

In recent years the observations revealed that in all types of
reactions large nonstatistical fluctuations exist
in the local particle density~[22]. Varying
the resolving power, i.e. the size of the phase-space domain in
which the particles are counted the density irregularities
resembled the scale-invariant, intermittent fluctuations that
appear in a turbulent fluid~[23-24]. This manifests itself in the
power-law  dependence of the normalized factorial moments on the
resolution characterized by the anomalous dimensions $d_q$ or
intermittency exponents $(q-1)d_q$. The factorial moments, $\xi_q$,
are defined by the Taylor expansion of ${\cal G}(z)$:
$$
    {\cal G}(z)=\sum_{q=0}^\infty{(z-1)^q\over q!}\,\xi_q.\eqno(3.1)
$$
Measuring factorial moments in a certain phase-space domain
is equivalent to measuring the
underlying multiplicity distribution. Let us consider the
relation between the $\xi_q$ and the $P_n$. The factorial moments
are expressed in terms of probabilities according to
$$
    \xi_q=q!\sum_{n=q}^\infty{n\choose q}\,P_n\eqno(3.2)
$$
where the summation produces the binomial moments. The count
probabilities take the form
$$
    P_n=\sum_{q=n}^\infty{q\choose n}\,{(-1)^{q-n}\over q!}
    \,\xi_q\eqno(3.3)
$$
when expressed in terms of factorial moments.
Comparing eqs. (3.2-3) to eqs. \hbox{(2.9-10)} we see that the
relationship between probabilities and factorial moments, generated by
${\cal G}(z)$ and ${\cal G}(1+z)$,
is the {\it same\/} as the relationship between
combinants and factorial cumulants, generated by
$\ln {\cal G}(z)$ and $\ln {\cal G}(1+z)$. This important feature
which is valid for ordinary moments and cumulants too was first
emphasized by Kauffmann and Gyulassy~[2].

The above correspondence
enables us to apply the same scaling argument for count
probabilities which was utilized in the previous section for
combinants. Let us assume that instead of the normalized factorial
cumulants, $K_q$, the normalized factorial moments
$F_q=\xi_q/\bar n^q$
obey the LPA-type recurrence relation with constant coefficients $A_q$:
$$
    F_q=A_qF_2^{q-1}.\eqno(3.4)
$$
Eq.~(3.4) is a special case of the famous Ochs-Wosiek empirical
relation~[25] that can be formulated as
$$
    F_q=A_qF_2^{(q-1)d_q/d_2}.\eqno(3.5)
$$
{}From currently available data for the factorial moments,
$A_q\approx1$ and the intermittency exponent
ratios $(q-1)d_q/d_2$ are largely independent of reaction type, energy
and phase-space dimension~[22].
If $F_2$ changes according to a power-law with varying resolution
eq.~(3.4) corresponds to  monofractal density fluctuations
characterized by $d_q/d_2=1$, i.e. by the unique anomalous
dimension $d_2$ given by the intermittency exponent of $F_2$.
Plugging into eq.~(3.3) with $\xi_q$ expressed according to
eq.~(3.4) and evaluating the ratios $P_n/\bar n$ one arrives at
$$
    {P_n\over\bar n}=
    \sum_{q=n}^\infty{q\choose n}{(-1)^{q-n}\over q!}
    A_q\, (\bar nF_2)^{q-1}=\eta_n(\bar nF_2).\eqno(3.6)
$$
That is to say, if eq.~(3.4) holds for the factorial moments
the $P_n/\bar n$ chosen at fixed $n$ not depend on
reaction type, energy, binsize and dimension arbitrarily, but
only through the momentum combination $\bar nF_2$.
Picking up the $P_n$ from the tail of the multiplicity distributions
one subtracts the contribution of the low-order
factorial moments from the scaling functions. In this manner we are
able to go further in testing the monofractality of
density fluctuations than the highest order
factorial moment that can be extracted from observational data.
Since monofractal patterns in multihadron final states
are expected to form during
a second-order QCD phase transition the collapse of the
$P_n/\bar n$ to the scaling curves $\eta_n(\bar nF_2)$
as $d_q/d_2\to1$ may serve as a signature of Quark-Gluon Plasma
formation.
This point will be discussed in the next section.
We note that the absence of translation invariance can lead to the
violation of the binsize-independence of $A_q$ and thus the scaling
behaviour of $P_n$.

\vfill\eject

\noindent
{\bo 4. Scaling laws in the Poisson cluster model}

Let us assume that the observed events are composed of identical groups
of particles (clusters, clans, Quark-Gluon Plasma droplets or whatever)
distributed according to a Poisson
process. The generating function of the total event
multiplicity distribution $P_n$ becomes~[26-27]
$$
    {\cal G}(z)=\exp\left(\bar{\cal C}({\cal H}(z)-1)\right)\eqno(4.1)
$$
which is the convolution of the Poissonian generating function
of the distribution of clusters having mean $\bar{\cal C}$ and the
generating function ${\cal H}(z)$ of the distribution of particles
within a single cluster.
Eq.~(4.1) is another way of writing the generating function of
infinitely divisible distributions, eq.~(2.5), with
$\bar{\cal C}=-\ln {\cal G}(0)$ and
$$
    {\cal H}(z)=
    1-{\ln{\cal G}(z)\over\ln {\cal G}(0)}=
    \sum_{q=1}^\infty p_qz^q.\eqno(4.2)
$$
In eq.~(4.2) the $q$-particle count probability in a cluster, $p_q$,
is found to be~[10-11]
$$
    p_q={C_q\over\sum_q C_q}.\eqno(4.3)
$$
Since ${\cal H}(0)=0$ each cluster must contain at least one particle,
i.e. $p_0=C_0=0$.
The two basic parameters of the model, the average
cluster multiplicity, $\bar{\cal C}$, and the average multiplicity
in a cluster, $\bar q$, are given by~[28,12]
$$
    \bar{\cal C}=-\ln P_0\qquad\hbox{\rm and}\qquad
    \bar q=\bar n/\bar{\cal C}.\eqno(4.4)
$$
In the Poisson cluster picture
the normalized factorial moments of a single cluster, ${\cal F}_q$,
can be expressed in terms of $\bar{\cal C}$ and the normalized
factorial cumulants $K_q$ of the total events as~[29-30]
$$
    {\cal F}_q=\bar{\cal C}^{q-1}K_q.\eqno(4.5)
$$
Writing $K_q$ according to the LPA relation, eq.~(2.11),
yields~[28,11]
$$
    {\cal F}_q=A_q{\cal F}_2^{q-1}.\eqno(4.6)
$$
Thus we see that a Poisson superposition of clusters which obey the
Ochs-Wosiek relation, eq.~(3.5), with
intermittency exponent ratios $(q-1)$ leads to
the validity of the LPA relation, eq.~(2.11), for the total events.
Comparing eqs.~(2.3) and \hbox{(4.3-5)} one finds the following
relationship between the scaling functions derived in the
previous sections:
$$
    \chi_q(\bar nK_2)\,\big|_{\,\rm total\ event}
    =\,
    \eta_q(\bar nF_2)\,\big|_{\,\rm single\ cluster}.
    \eqno(4.7)
$$
The two sets of scaling functions appearing
on the two different levels are equivalent.
On the basis of eqs.~(4.5-7) let us collect the correspondences
between total event observables and quantities characterizing a
single cluster:

\item{\it i)\/}    If the events are Poisson superpositions of
{\it selfsimilar\/} clusters we observe the power-law dependence of
$K_q$ instead of $F_q$ (the $F_q$ show a bending upward behaviour on
log-log plot). Through the slopes of the $K_q$ we actually see
the slopes of the ${\cal F}_q$, that is, the fractal structure of a
single cluster~[29-30].

\item{\it ii)\/}   When the Poisson clusters underlying the events
have {\it monofractal\/}
structure and the ${\cal F}_q$ obey eq.~(3.4) the $K_q$ exhibit
the LPA relation, eq.~(2.11). Thus the validity of LPA may arise as
the consequence of randomly superimposed monofractal clusters~[28,11].

\item{\it iii)\/}  Accordingly, by the observation of $\chi_q$-scaling
for the total events confirming the validity of LPA one actually
sees $\eta_q$-scaling for single clusters confirming their monofractal
structure.

\noindent
The search of monofractal patterns in multihadron final states plays
a distinguished role in intermittency analyses. It has been conjectured
that monofractal density fluctuations could signal a phase transition
from the Quark-Gluon Plasma [31-32]. At the critical point of a
second-order
QGP phase transition to the hadron gas intermittent fluctuations
are expected to occur characterized by a unique anomalous dimension
as in the Ising model~[33].
Bia\l as and Hwa proposed the experimental confirmation of the validity
of eq.~(3.4) as a QGP signature, i.e. the $q$-independence of the
anomalous dimensions $d_q$ in eq.~(3.5)~[31]. Another proposal, put
forward by Peschanski, is based on the description of the monofractal
fluctuations at the critical point in terms of clusters~[32]. The unique
anomalous dimension is interpreted in this case as the fractal dimension
of the random set of intermittent clusters of the ordered phase inside
the disordered one at the transition. For each proposal one has to
assume that the monofractal density fluctuations
survive the further evolution
of the system toward the hadronic final state. According to eq.~(4.7)
both sets of scaling functions could signal QGP formation.
Events possessing homogeneous monofractal fluctuations exhibit
$\eta_q$-scaling whereas events composed of random monofractal clusters
give rise (through the $\eta_q$-scaling of the clusters) to
$\chi_q$-scaling for the observed events.

\noindent
{\bo 5. Conclusions}

Over the last few years the analysis of fluctuations and correlations
in restricted domains of phase-space revealed that the higher-order
statistics characterizing the fluctuation pattern are frequently
related to two-particle statistics in a very simple manner~[22].
One example is the LPA relation, eq.~(2.11), for the factorial cumulant
moments. The validity of LPA was confirmed for $p\bar p$ collisions
at CERN collider energies with constant coefficients $A_q$ close in
magnitude to the Negative Binomial values $A_q=(q-1)!$~[18].
Another frequently encountered relationship between second- and
higher-order statistics is the Ochs-Wosiek relation for the factorial
moments. In addition to the simplicity of eq.~(3.5) it has a kind of
universality: rather independently of reaction type, energy,
binsize and phase-space dimension,
the coefficients $A_q\approx1$ and the intermittency exponent
ratios \hbox{$(q-1)d_q/d_2$} are compatible with a
L\'evy-stable law with L\'evy-index $\mu=1.6$~[22].

Eq.~(3.4), the Ochs-Wosiek relation with intermittency exponent ratios
\hbox{$(q-1)$}
characterizes monofractal density fluctuations. Such patterns
are expected to form during a second-order phase transition from the
Quark-Gluon Plasma to the hadron gas [31-32]. Recent studies of the
phase structure of QCD suggest that for two massless quark flavours
QCD undergoes a second-order transition and this remains valid
for three flavours provided
that the strange quark mass is sufficiently large~[34]. Hence it is of
interest to find clear signatures of monofractal density
fluctuations appearing in multihadron final states.

In this paper we have derived a scaling law for the
count probabilities $P_n$ in the presence of monofractal density
fluctuations. It is expressed by eq.~(3.6). According to this scaling
law, if one plots the ratios $P_n/\bar n$ chosen at fixed $n$ against
the momentum combination $\bar nF_2$ the validity of eq.~(3.4)
with constant coefficients $A_q$ results in a universal curve,
$\eta_n(\bar nF_2)$, instead of many
different behaviours corresponding to different reaction types,
energies, binsizes and phase-space dimensions.
Increasing $n$ the low-order factorial
moments can be excluded in a systematic manner from testing eq.~(3.4).
We have also found that a similar scaling law is satisfied for
monofractal density fluctuations occuring in randomly distributed
clusters. In this case
the combinants show up a scaling behaviour: at fixed $n$ the ratios
$C_n/\bar n$ fall onto a universal curve when plotted against the
momentum combination $\bar nK_2$. Through the scaling curves one
observes the validity of eqs.~(3.4) and (3.6)
again but now for a single cluster. Both types of scaling rules provide
easily measurable, clear signatures of
homogeneous and clustered monofractal density fluctuations
which could be the remnants of a second-order QCD phase transition.
It will be interesting to see how these scaling predictions
are satisfied in next heavy ion experiments.

\noindent
{\bo Acknowledgements}

I would like to express my thanks to W. Kittel  for a stimulating conversation.
This work was supported by the Hungarian Science Foundation
under grants No. OTKA-2972 and OTKA-F4019.

\vfill\eject

\parskip=0pt\hsize=15.5truecm
\noindent
{\bo References}
\vskip.5truecm
\frenchspacing

\item{[1]}  M. Gyulassy and S.K. Kauffmann, {\it Phys. Rev. Lett.\/}
            \ub{40} (1978) 298.\ss
\item{[2]}  S.K. Kauffmann and M. Gyulassy, {\it J. Phys. A\/}
            \ub{11} (1978) 1715.\ss
\item{[3]}  J. Bartke, {\it Phys. Scripta\/}
            \ub{27} (1983) 225.\ss
\item{[4]}  P. Carruthers and C.C. Shih, {\it Int. J. Mod. Phys. A\/}
            \ub{2} (1987) 1447.\ss
\item{[5]}  A.B. Balantekin and J.E. Seger, {\it Phys. Lett. B\/}
            \ub{266} (1991) 231.\ss
\item{[6]}  A.J. Buras, {\it Nucl. Phys. B\/}
            \ub{56} (1973) 275.\ss
\item{[7]}  S. Hegyi and S. Krasznovszky,
            {\it Phys. Lett. B\/} \ub{251} (1990) 197.\ss
\item{[8]}  M. Charlet, in {\it Fluctuations and Fractal
            Structure,\/}\hfill\break
            Proc. of the Ringberg Workshop on Multiparticle Production, p.~140,
            \hfill\break ed. by R.C. Hwa, W. Ochs and N. Schmitz, World
            Scientific (1992).\ss
\item{[9]}  B.A. Li,
            {\it Phys. Lett. B\/} \ub{300} (1993) 14.\ss
\item{[10]} I. Szapudi and A.S. Szalay,
            {\it Astrophys. J.\/} \ub{408} (1993) 43.\ss
\item{[11]} S. Hegyi,
            {\it Phys. Lett. B\/} \ub{309} (1993) 443.\ss
\item{[12]} S. Lupia, A. Giovannini and R. Ugoccioni,
            {\it Z. Phys. C\/} \ub{59} (1993) 427.\ss
\item{[13]} W. Feller, {\it Introduction to Probability Theory and
            its Applications, Vol. I.\/}, \hfill\break
            Third Edition, Wiley (1971).\ss
\item{[14]} P. Carruthers, {\it Phys. Rev. A\/}
            \ub{43} (1991) 2632.\ss
\item{[15]}  P. Carruthers, {\it Acta Phys. Pol. B\/}
            \ub{22} (1991) 931.\ss
\item{[16]}  P. Carruthers and I. Sarcevic, {\it Phys. Rev. Lett.\/}
            \ub{63} (1989) 1562.\ss
\item{[17]} P. Carruthers, H.C. Eggers, Q. Gao and I. Sarcevic,
            {\it Int. J. Mod. Phys. A\/} \hfill\break
            \ub{6} (1991) 3031.\ss
\item{[18]} P. Carruthers, H.C. Eggers and I. Sarcevic,
            {\it Phys. Lett. B\/} \ub{254} (1991) 258.\ss
\item{[19]} EHS/NA22 Collaboration, N. Agababyan et al.,
            {\it Z. Phys. C\/} \ub{59} (1993) 405.\ss
\item{[20]} H.C. Eggers, P. Carruthers, P. Lipa and I. Sarcevic,
            {\it Phys. Rev. C\/} \hfill\break
            \ub{44} (1991) 1975.\ss
\item{[21]} E.A. De Wolf, in {\it Fluctuations and Fractal
            Structure,\/}\hfill\break
            Proc. of the Ringberg Workshop on Multiparticle Production,
            p.~222, \hfill\break ed. by R.C. Hwa, W. Ochs and N. Schmitz,
            World Scientific (1992).\ss
\item{[22]} E.A. De Wolf, I. Dremin and W. Kittel,
            {\it Scaling Laws for Density \hfill\break Correlations
            and Fluctuations in Multiparticle Dynamics,\/} \hfill\break
            Nijmegen University Report HEN-362/1993,
            to appear in Physics Reports.\ss
\item{[23]} A. Bia\l as and  R. Peschanski, {\it Nucl. Phys. B\/}
            \ub{273} (1986) 703.\ss
\item{[24]} A. Bia\l as and R. Peschanski, {\it Nucl. Phys. B\/}
            \ub{308} (1988) 857.\ss
\item{[25]} W. Ochs and J. Wosiek,
            {\it Phys. Lett. B\/} \ub{214} (1988) 617.\ss
\item{[26]} A. Giovannini and L. Van Hove, {\it Acta Phys. Pol. B\/}
            \ub{19} (1988) 495.\ss
\item{[27]} J. Finkelstein, {\it Phys. Rev. D\/}
            \ub{37} (1989) 2446.\ss
\item{[28]} S. Hegyi and T. Cs\"org\H o, {\it Phys. Lett. B\/}
            \ub{296} (1992) 256.\ss
\item{[29]} H.C. Eggers, Ph.D. Thesis, University of Arizona (1991).\ss
\item{[30]} P. Lipa, in {\it Fluctuations and Fractal Structure,\/}\hfill\break
            Proc. of the Ringberg Workshop on Multiparticle Production, p.~96,
            \hfill\break ed. by R.C. Hwa, W. Ochs and N. Schmitz, World
            Scientific (1992).\ss
\item{[31]} A. Bia\l as and R.C. Hwa, {\it Phys. Lett. B\/}
            \ub{253} (1991) 436.\ss
\item{[32]} R. Peschanski, CERN Report, CERN-TH/90-5963, \hfill\break
            Presented at the Strasbourg Workshop on Quark-Gluon Plasma
            Signatures, \hfill\break October 1990.
\item{[33]} H. Satz, {\it Nucl. Phys. B\/}
            \ub{326} (1989) 613.\ss
\item{[34]} F. Wilczek, {\it Remarks on Hot QCD,\/}
            Princeton University Report \hfill\break IASSNS-HEP-93/48,
            Invited talk given at Quark Matter '93.

\bye